\documentclass[letter]{aa}

\usepackage{txfonts}

\usepackage[T1]{fontenc}
\usepackage{float}
\usepackage[colorlinks=true,     linkcolor=blue, citecolor=blue, filecolor=blue, urlcolor=blue]{hyperref}

\DeclareRobustCommand{\VAN}[3]{#2}
\let\VANthebibliography\thebibliography
\def\thebibliography{\DeclareRobustCommand{\VAN}[3]{##3}\VANthebibliography}

\usepackage{graphicx}	
\usepackage{amsmath}	
\usepackage{amssymb}	
\usepackage{gensymb}
\usepackage[
singlelinecheck=false 
]{caption}
\usepackage{threeparttable}
\usepackage{float}
\usepackage{upgreek}
\usepackage{color}
\definecolor{darkgreen}{rgb}{0,0.5,0}
\definecolor{purple}{rgb}{1,0,1}
\newcommand{\kibitz}[2]{\ifnum\Comments=1\textcolor{#1}{#2}\fi}

\newcommand{\per}{$^{-1}$}

\begin{document}

\title{The MeerKAT Massive Distant Clusters Survey: A Radio Halo in a Massive Galaxy Cluster at \textit{z} = 1.23}

\author{S. P. Sikhosana
        \inst{1,2}\thanks{E-mail:sikhosanas@ukzn.ac.za}
        \and
        M. Hilton
        \inst{3,2}
        \and
        G. Bernardi
        \inst{4,5,6}
        \and
        K. Kesebonye
        \inst{3}
        \and
        D. Y. Klutse
        \inst{1,2}
        \and
        K. Knowles
        \inst{1,5,6}
        \and
        K. Moodley
        \inst{1,2}
        \and
        T.~Mroczkowski
        \inst{7}
        \and
        B. Partridge
        \inst{8}
        \and
        C. Sifón
        \inst{9}
        \and
        C. Vargas
        \inst{10}
        \and
        E. Wollack
        \inst{11}
       }

   \institute{
   Astrophysics Research Centre, University of KwaZulu-Natal, Durban, 3696, South Africa
   \and
   School of Mathematics, Statistics $\&$ Computer Science, University of KwaZulu-Natal, Westville Campus, Durban 4041, South Africa
   \and
   Wits Centre for Astrophysics, School of Physics, University of the Witwatersrand, Private Bag 3, 2050, Johannesburg, South Africa
   \and
   INAF-Istituto di Radioastronomia, via Gobetti 101, 40129 Bologna, Italy
   \and
   Centre for Radio Astronomy Techniques and Technologies, Department of Physics and Electronics, Rhodes University, P.O. Box 94, Makhanda 6140, South Africa
   \and
   South African Radio Astronomy Observatory, 2 Fir Street, Black River Park, Observatory, Cape Town 7925, South Africa
   \and
   European Southern Observatory, Karl-Schwarzschild-Str. 2, 85748 Garching, Germany
   \and
   Department of Astronomy, Haverford College, Haverford, PA 19041, USA
   \and
   Instituto de Física, Pontificia Universidad Católica de Valparaíso, Casilla 4059, Valparaíso, Chile
   \and
    Instituto de Astrof\'isica and Centro de AstroIngenier\'ia, Facultad de F\'isica, Pontificia Universidad Cat\'olica de Chile, Av. Vicu\~na Mackenna 4860, 7820436 Macul, Santiago, Chile 
   \and
   NASA Goddard Space Flight Center, 8800 Greenbelt Rd, Greenbelt, MD 20771, USA
}

\date{Accepted XXX. Received YYY; in original form ZZZ}

\abstract{
In the current paradigm, high redshift radio halos are expected to be scarce due to inverse Compton energy losses and redshift dimming, which cause them to be intrinsically faint. This low occurrence fraction is predicted by cosmic ray electron turbulent re-acceleration models. To date, only a handful of radio halos have been detected at redshift \textit{z} > 0.8. We report the MeerKAT detection of a radio halo hosted by a galaxy cluster ACT-CL\,J0329.2-2330 at $z= 1.23$, making it the highest redshift halo detected thus far. Using $L$-band and $UHF$-band observations, we derive a radio halo spectral index of $ \alpha^{1.3GHz}_{0.8GHz}$ = 1.3 $\pm$ 0.4 and a radio power of P$_\mathrm{1.4~GHz} = (4.4 \pm 1.5) \times 10^{24}$\,W\,Hz$^{-1}$. This result further confirms that there is rapid magnetic field amplification in galaxy clusters at high redshift.} 


\maketitle



\section{Introduction}
Galaxy clusters, which are at the top of the hierarchical structure formation, grow by accreting smaller systems during cluster mergers \citep{1974ApJ...187..425P}. These merger events result in the release of enormous amounts of energy ($\sim$10$^{64}$\,ergs), which is partially dispersed through shocks and turbulence in the intracluster medium (ICM). The turbulent activity induces magnetic field amplification and the (re)acceleration of cosmic ray electrons (CRes) to relativistic energies ($\gamma \sim 10^3$ -- $10^4$). Such environments are conducive to the formation of cluster-wide synchrotron radiation that is not attributed to individual galaxies (like in the case of emission from active galactic nuclei). This Mpc-scale non-thermal diffuse emission is generally categorised as radio halos and relics. The origin of halos and relics has long been debated by proponents of Leptonic and Hadronic models, with observations favouring Leptonic models \citep[see][for reviews]{2014IJMPD..2330007B,2019SSRv..215...16V}.
\par
Radio relics are highly polarised (up to $\sim$70$\%$) \citep{2022A&A...666A...8S,2023A&A...675A..51D}, elongated radio sources located at the periphery of galaxy clusters. Recent observations indicate that relics also exhibit filamentary morphologies \citep{2020A&A...636A..30R, 2022A&A...659A.146D}. The origin of relics is linked to the shock waves that propagate in the ICM during cluster mergers. The electrons scattering upstream and downstream of the shock region are accelerated to relativistic energies via diffusive shock acceleration \citep[DSA;][]{1949PhRv...75.1169F,1983RPPh...46..973D,2007MNRAS.375...77H}. X-ray studies reveal that the Mach numbers of cluster shocks are too low ($\mathcal{M}$ $<$ 3) to reproduce the observed relic structures and luminosities purely from accelerating electrons in the thermal pool \citep{2005ApJ...627..733M,2016MNRAS.460L..84B,2019ApJ...873...64D}. Thus, the current scenario is that relics are formed by the DSA of a pre-accelerated CRe population \citep{2013MNRAS.435.1061P, 2014ApJ...785....1B,2020A&A...634A..64B}. Indeed, a few recent studies have connected relics to head-tail radio galaxies, which are candidate sources of the seed electrons \citep[e.g.][]{2017ApJ...835..197V,2021A&A...646A..56R}.
\par
Radio halos are located in the central region of galaxy clusters and have a regular morphology, which tends to trace the X-ray emitting ICM \citep{2001A&A...369..441G}. Unlike relics, halos have low polarisation percentages. From current observations, the origin of halos strongly aligns with the Leptonic model, where pre-existing seed relativistic electrons are re-accelerated by merger-induced turbulence via stochastic Fermi II-type processes \citep{2001MNRAS.320..365B,2005MNRAS.357.1313C,2012A&ARv..20...54F,2014MNRAS.443.3564D}. Observations have linked radio halos to host clusters with high masses (M$_{500}$ > 4$\times 10^{14}$M$_{\odot}$) and X-ray and/or optical merger signatures \citep{2005ApJ...627..733M,2013ApJ...777..141C,2014MNRAS.440.2901S,2014ApJ...786...49L,2015A&A...579A..92K,2019MNRAS.486.1332K,2021MNRAS.500.2236R}. The presence of radio halos with ultra-steep radio spectra\footnote{ S$_{\nu}$ $\propto$ $\nu^{-\alpha}$, where S$_{\nu}$ is the source flux density at frequency $\nu$ and $\alpha$ is the spectral index used in this work.} ($\alpha$ > 1.5) further supports the Leptonic model. It has been theoretically shown that the re-acceleration of secondary electrons arising from proton-proton collisions may also contribute to synchrotron radiation \citep{2011MNRAS.410..127B,2017MNRAS.465.4800P}, even with the $\gamma$-ray limits of the Fermi-LAT observations \citep{2021A&A...648A..60A}. However, this will not be a significant contribution, unlike the case for Hadronic models \citep{1980BAAS...12..471D,2000A&A...362..151D}. The sensitivity of low-frequency instruments, such as the Low-Frequency Array \citep[LOFAR;][]{2013A&A...556A...2V}, has resulted in the discovery of new classes of diffuse emission. The recent discovery of mega halos \citep{2022Natur.609..911C} has further challenged our understanding of CRe transportation in galaxy clusters. 

\par
At high redshifts, the surface brightness of radio halos and relics is affected by the inverse Compton (IC) energy losses of the CRe (CRe IC scattering lifetimes scale as (1 + z)$^{4}$) and redshift dimming effects \citep{2014IJMPD..2330007B,2021NatAs...5..268D}. It is therefore expected that the radio halo occurrence rate decreases at higher redshifts. \citet{2014ApJ...786...49L} were the first to detect a radio halo at a relatively high redshift of $z=0.870$. The host cluster is the massive ACT-CL\,J0102-4915 (`El Gordo'; \citealt{Menanteau_2012}, $M_{\rm 500c}=(1.17 \pm 0.17) \times 10^{15}$\, M$_{\odot}$). For some time, this had been the only radio halo detected at $z \gtrsim 0.8$. \citet{2021A&A...650A.153D} reported a tentative detection of diffuse emission in SPT\,J2016--5844, which is at redshift $z$= 1.13. However, they could not claim a definite detection due to the limited resolution of the Australian Square Kilometre Array Pathfinder and the lack of sensitivity of the Australia Telescope Compact Array.
\par
Due to the scarcity of high redshift halos and relics, statistical studies have been constrained to nearby galaxy cluster samples \citep[$z \sim 0.1 - 0.4$;][]{2013ApJ...777..141C,2021A&A...647A..51C}. Recently, \citet{2021NatAs...5..268D} observed a \textit{Planck}$-$selected sample of high redshift (0.6 < $z$ < 0.9) clusters \citep{2016A&A...594A..27P}. In their study, they reported a halo occurrence fraction of 50$\%$ and that the halos exhibited radio luminosities similar to those hosted by nearby clusters ($z \sim 0.2$). They hence concluded that observing such luminosities at higher redshifts would mean that magnetic fields are amplified faster during the first phases of cluster formation. Therefore, this implies that the magnetic field strengths in high redshift clusters are similar to those in galaxy clusters at lower redshift.
\par
In this letter, we report the discovery of the highest-redshift radio halo now known, detected in deep MeerKAT $L$ and $UHF$-band observations, hosted by ACT-CL\,J0329.2-2330 (herein ACT-CL\,J0329). ACT-CL\,J0329 is reported in the SZ cluster catalogue compiled as part of the Atacama Cosmology Telescope (ACT) Data Release 5 \citep{2021ApJS..253....3H}, and also appears in the SPTpol Extended Cluster Survey catalogue \citep{Bleem_2020}. It is at a redshift $z$ = 1.23 and has a mass of M$_{\rm 500c} = 9.7^{+1.7}_{-1.6}\times 10^{14}$\,M$_{\odot}$ \citep{2021ApJS..253....3H}. The ACT SZ cluster properties are summarised in Table \ref{tab:j0329}. This radio halo detection gives us insight into non-thermal astrophysical phenomena that occurred when the universe was less than half its current age ($\sim$8.5 billion years ago). The letter is organised as follows. We outline the MeerKAT observations and data reduction in Section \ref{sec:observe}. In Section \ref{sec:results}, we present the MeerKAT results, and in Section \ref{sec:conclude}, we present our conclusions. We adopt a $\Lambda$CDM flat cosmology with $H_0 = 70$\,km\,s{\per}\,Mpc{\per}, $\Omega_m = 0.3$, and $\Omega_\Lambda = 0.7$. At the redshift of ACT-CL\,J0329 ($z$ = 1.23), the luminosity distance is 8535.6\,Mpc, and 1$\arcsec$ corresponds to 8.32\,kpc.

\begin{table}[]
	\centering
	\caption{Properties of ACT-CL\,J0329.2-2330.}
	\label{tab:bullet}
	\begin{tabular}{lc} 
		\hline
		R.A.$_{\rm J2000}$ (hh:mm:ss.s) & 03:29:16.54  \\
		Dec.$_{\rm J2000}$ (dd:mm:ss.s) & $-$23:30:07.56 \\
		redshift & 1.23 \\
		$M_{\rm 500cCal,SZ} \; (10^{14}\;M_\odot)$ & 9.7$^{+1.7}_{-1.6}$ \\
		$M_{\rm 500cUncorr,SZ} \; (10^{14}\;M_\odot)$ & 8.0$^{+1.3}_{-1.1}$ \\
        Signal-to-Noise ratio  & 22.1\\
		\hline
        \label{tab:j0329}
	\end{tabular}
 \caption*{\textbf{Notes}. SZ values are from the ACT DR5 catalogue \citep{2021ApJS..253....3H}. $M_{\rm 500cCal,SZ}$ is the ACT richness-based weak-lensing calibrated cluster mass (thought to be the best estimate of mass). $M_{\rm 500cUncorr,SZ}$ is the ACT cluster mass estimate assuming the \citet{Arnaud_2010} scaling relation, without Eddington bias correction, which is used for comparison with masses in the \textit{Planck} PSZ2 cluster catalogue.}
\end{table}

\section{MeerKAT Observations and Data Reduction}
\label{sec:observe}
MeerKAT is a 64-antenna array operated by the South African Radio Astronomy Observatory \citep[SARAO;][]{2009IEEEP..97.1522J,2018ApJ...856..180C}. MeerKAT's longest baseline is 8\,km, resulting in 6$\arcsec$ resolution at $L$-band (900 $-$ 1670\,MHz). Approximately three-quarters of the antennas are located within a 1\,km radius, with a minimum baseline of 29\,m resulting in a $UHF$-band maximum scale of 52$\arcmin$, making MeerKAT an ideal instrument for observing faint extended emission. 

ACT-CL\,J0329 was initially observed as part of the MeerKAT Massive Distant Cluster Survey (MMDCS) project proposal. This survey ultimately aims to obtain $L/UHF$-band observations of the 30 most massive clusters at $z$ > 1 in the ACT
DR5 catalogue \citep{2021ApJS..253....3H} with the purpose of studying diffuse radio emission, star formation, and AGN activity. The selected clusters also fall within the Dark Energy Camera Legacy Survey \citep[DECaLS;][]{Dey_2019} optical/infrared survey footprint. The MeerKAT observations were carried out at $L$-band with a total on-target time of 3.5 hours, using a dump rate of 8 seconds and 4096 channels. J0408$-$6545 and J0409$-$1757 were observed for the amplitude and phase calibration. The detection of a radio halo in ACT-CL\,J0329, which was part of the sample, prompted the submission of a follow-up proposal at $UHF$-band (580 $-$ 1015 MHz) (PI: S.P.\ Sikhosana). For the $UHF$ observations, the target was observed for 5 hours using the 4k channel mode, and the dump rate was 4 seconds. A summary of the observations is in Table \ref{tab:observations}. We detail the methods used to calibrate and image the data below.   

\par
For both $L$ and $UHF$-band observations, we used the Containerized Automated Radio Astronomy Calibration (CARACal\footnote{\url{https://github.com/caracal-pipeline/caracal}}) pipeline \citep{2020ascl.soft06014J} to calibrate and image the raw visibilities. CARACal uses the Stimela\footnote{\url{https://github.com/ratt-ru/Stimela-classic}} \citep{makhathini2018} framework to execute tasks from multiple astronomy software tools. The pipeline follows the standard radio interferometric data reduction steps, which include flagging, primary calibration, self-calibration, and imaging. We started by flagging all the frequency bands known to be affected by radio frequency interference (RFI) using the Common Astronomy Software Application \citet[\textsc{CASA};][]{2022PASP..134k4501C}\footnote{\url{https://casa.nrao.edu/}} \textsc{flagdata} task, these flags account for the high flagged percentages reported in Table \ref{tab:observations}. We used \textsc{Tricolour} \citep{2022ASPC..532..541H} to excise the remaining low-level RFI. We used the \textsc{casa} tasks for producing delay, bandpass, and gain calibration solutions for the primary calibration step. We used \textsc{setjy} to correct for the absolute flux density scale of J0408$-$6545. Self-calibration was carried out using \textsc{CubiCal} \citep{2018MNRAS.478.2399K}. The imaging was done using \textsc{WSClean} \citep{2014MNRAS.444..606O} with multiscale and wideband deconvolution algorithms enabled and the \textsc{Briggs} \textsc{robust} \citep{1995PhDT.......238B} set to 0. The details of the full-resolution (FR) images are presented in Table \ref{tab:observations}, and the primary beam-corrected images are shown in Figures \ref{fig:l-fullfield} and \ref{fig:uhf-fullfield}.

\begin{table}  
    \centering
    \begin{threeparttable}
    \caption{MeerKAT $L$ and $UHF$-band observations and full resolution image properties.}
    \begin{tabular}{lrrrrccl}
\hline
& $L$ & $UHF$\\ \hline \hline
Central frequency (MHz) & 1283 & 815.9 &\\
$\Delta \nu$  before flagging (MHz) & 856 & 435 &\\
No. of antennas & 64 & 60\\
Observing date (Y-M-D) & 2023-01-27 & 2023-10-12\\
Amplitude calibrator & J0408$-$6545 & J0408$-$6545\\
Phase calibrator & J0409$-$1757  & J0409$-$1757\\
On-source time (hrs) & 3.5 & 5\\
$t_{\rm int}$ (sec) & 8 & 4\\
Flagged ($\%$) & 55 & 36\\

$\theta_{\rm synth}$ (\arcsec $\times$ \arcsec) & 8.0 $\times$ 6.7 & 10.5 $\times$ 8.6\\
p.a. (\degree) & 162.4 & 142.7\\
$\sigma_{\rm rms}$ ($\upmu$Jy/beam) & 6.6 & 13.2\\
\hline \hline    
\label{tab:observations}
\end{tabular}
\end{threeparttable}
\end{table}

\par
In Figure \ref{fig:halo}, faint diffuse emission is visible in the central region of the cluster in FR images; however, there are bright compact sources embedded. We then implemented further imaging steps to model the compact sources, remove them in the visibility plane, and image at low resolution. These steps were carried out for both $L$ and $UHF$-band datasets. We began by modelling the compact sources in the field by producing a high-resolution (HR) image using \textsc{WSclean's} \textsc{inner-tukey} taper set to 10\,k$\lambda$, which corresponds to 173\,kpc at the cluster redshift. The chosen taper sufficiently captures the compact sources while resolving the diffuse emission component. We used the model column generated by the HR images to subtract the compact sources in the visibility plane; this was done by the \textsc{casa} task \textsc{uvsub}. We then produced a FR compact-source-subtracted image and used it to quantify the contamination from the removal of compact sources (see Figures \ref{fig:l-subtracted} and \ref{fig:uhf-subtracted}). We used the method described in \citet{2021A&A...651A.115V} to quantify the contamination. Finally, we produced a low-resolution (LR) image by applying an outer taper of 12\,k$\lambda$. The flux density measurements of the diffuse radio sources were extracted from the LR images using polygon regions for each source, which are guided by the 3$\sigma$ contour levels. The uncertainty  associated with the flux density of the diffuse radio source, measured at frequency $\nu$ is given by
\begin{equation}
\Delta S_{\nu} = \sqrt{(\delta S_{\nu} \times S_\nu)^2 + N_{beams} \sigma_{rms}^2+\sigma_{sub}^2} \;,
\end{equation}
where $\delta S_{\nu}$ is the flux calibration uncertainty, which we assume to be 5$\%$ for $L$-band \citep{2022A&A...657A..56K} and 14$\%$ for $UHF$-band (Obtained from cross-matching with NVSS sources within the primary beam, using $\alpha$=0.7 to scale the flux densities.), $N_{beams}$ is the number of beams within the region that the flux density was measured, and $\sigma_{rms}$ is the local rms noise of the image. $\sigma_{sub}$ is the uncertainty due to compact-source-subtraction \citep{2021A&A...651A.115V}, it is given by
\begin{equation}
\sigma_{sub} = \sum_{i}N_{beam,s_{i}} \sigma_{rms}^2\;,
\end{equation}
where the sum is taken over the number of compact sources ($i$) subtracted within the polygon region of the diffuse radio source and $N_{beam,s_{i}}$ is the number of beams covering the $i^{th}$ compact source. 
\par 
We used UV- and resolution-matched images, which are centred at 1.28\,GHz and 815.9\,MHz, to create a spectral index map. To ensure that both images cover the same UV range at the shortest baselines, we applied a \textsc{minuv} cut of 100\,$\lambda$. We matched the resolution of the images by re-imaging the $L$-band data with beam parameters matching the $UHF$ LR image. The spectral index map was produced using Broadband Radio Astronomy Tools \citep[BRATS,][]{2013MNRAS.435.3353H,2015MNRAS.454.3403H}. We set the cutoff to 3$\sigma$ and the calibration uncertainty to 5$\%$; the rest of the parameters were left to default settings.

\section{Results and Discussion}
\label{sec:results} 
\begin{figure}
	\includegraphics[width=\columnwidth]{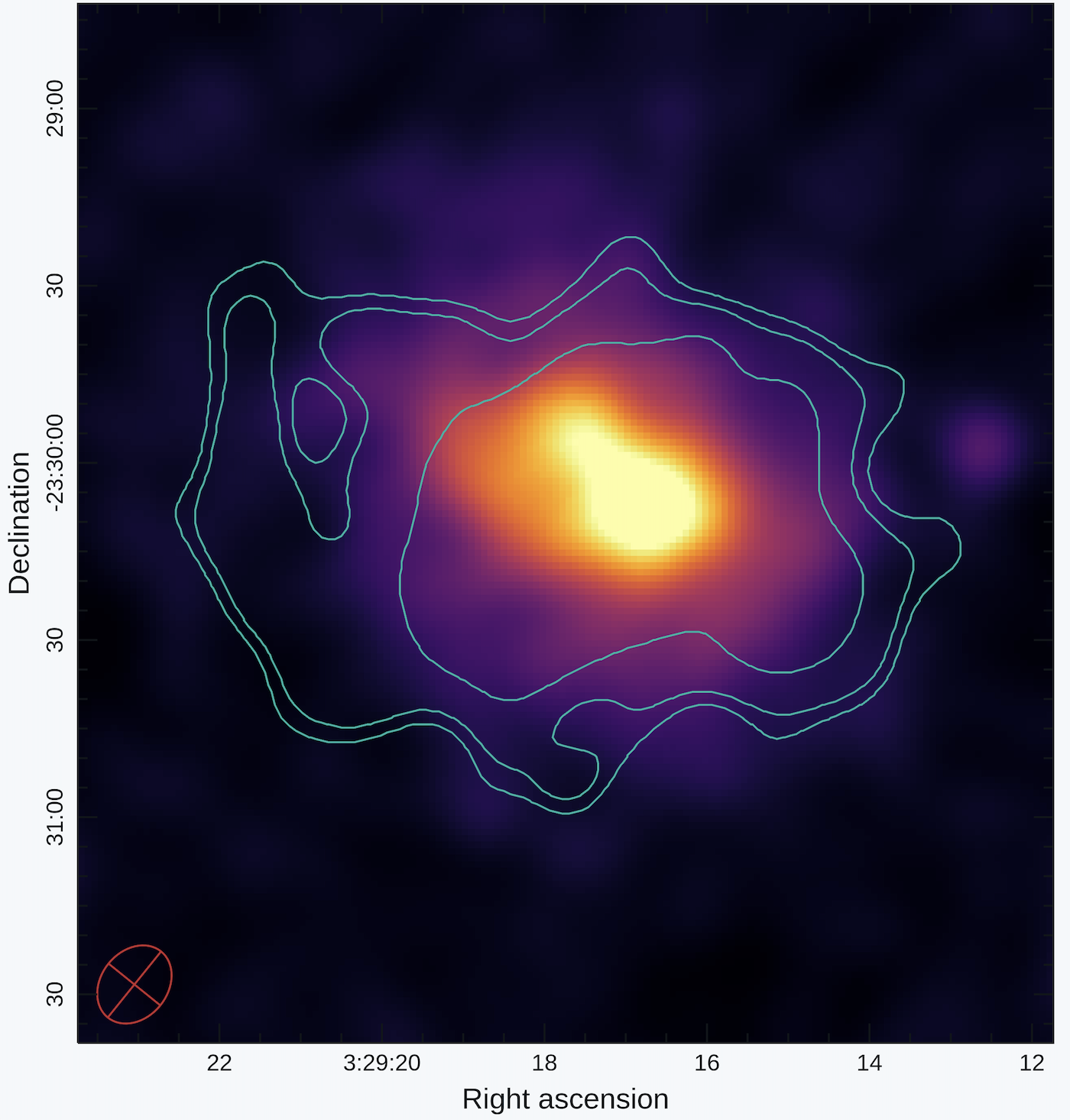}
    \caption{Archival \textit{Chandra} image of ACT-CL\,J0329, observation IDs: 18282 and 18882. The cluster was observed for 47 ksec, using ACIS-I in VFAINT mode, and the image covers the 0.5-4 keV energy band. The image was smoothed to a FWHM of 12$\arcsec$. The overlaid contours are from the MeerKAT $UHF$-band LR images; the contour levels are $\sigma \times$[3,4,8], where 1$\sigma$ = 17.6 $\upmu$Jy/beam. The beam size is indicated by the red ellipse. }
\label{fig:halo-xray}
\end{figure}

The MeerKAT $L$ and $UHF$-band LR images reveal extended emission at the centre of ACT-CL\,J0329. This emission has a largest linear size of 1.1\,Mpc at 1.28 GHz and\,has regular morphology. The morphology traces the X-ray emission as shown in Figure \ref{fig:halo-xray}, where the X-ray image is obtained from archival \textit{Chandra} ACIS-I observations (observation IDs: 18282 and 18882) with an exposure time of 47 ksec, using the  0.5-4 keV energy band. Thus, we categorise the source as a radio halo. There is a nearby elongated source to the east that connects to the halo emission; in the FR image, this source appears to be a head-tail (HT) radio galaxy. After subtracting the compact emission of the HT galaxy, there is residual emission in the LR images, which we measure to have flux densities of 0.29 $\pm$ 0.05 mJy and 0.78 $\pm$ 0.3 mJy at $L$ and $UHF$-band, respectively. However, the emission of the elongated source is too faint in the LR images to produce a spectral index map; hence, we cannot unambiguously conclude whether it is a radio relic. The presence of the HT galaxy suggests that the seed electrons may be deposited into the ICM as proposed by the re-acceleration model \citep{2024MNRAS.528..141L}.

\begin{figure*}
	\includegraphics[scale=0.4]{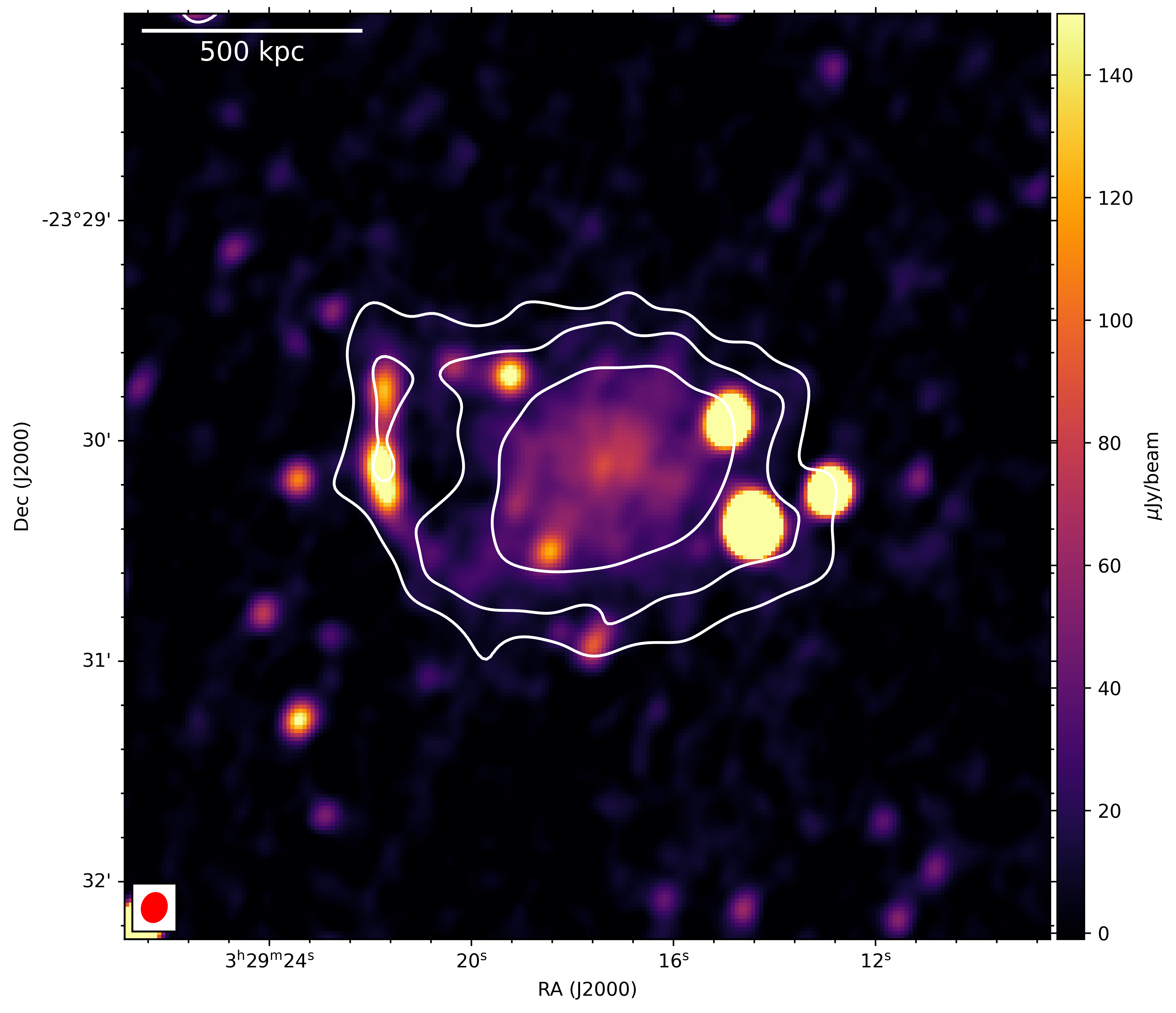}
    \includegraphics[scale=0.4]{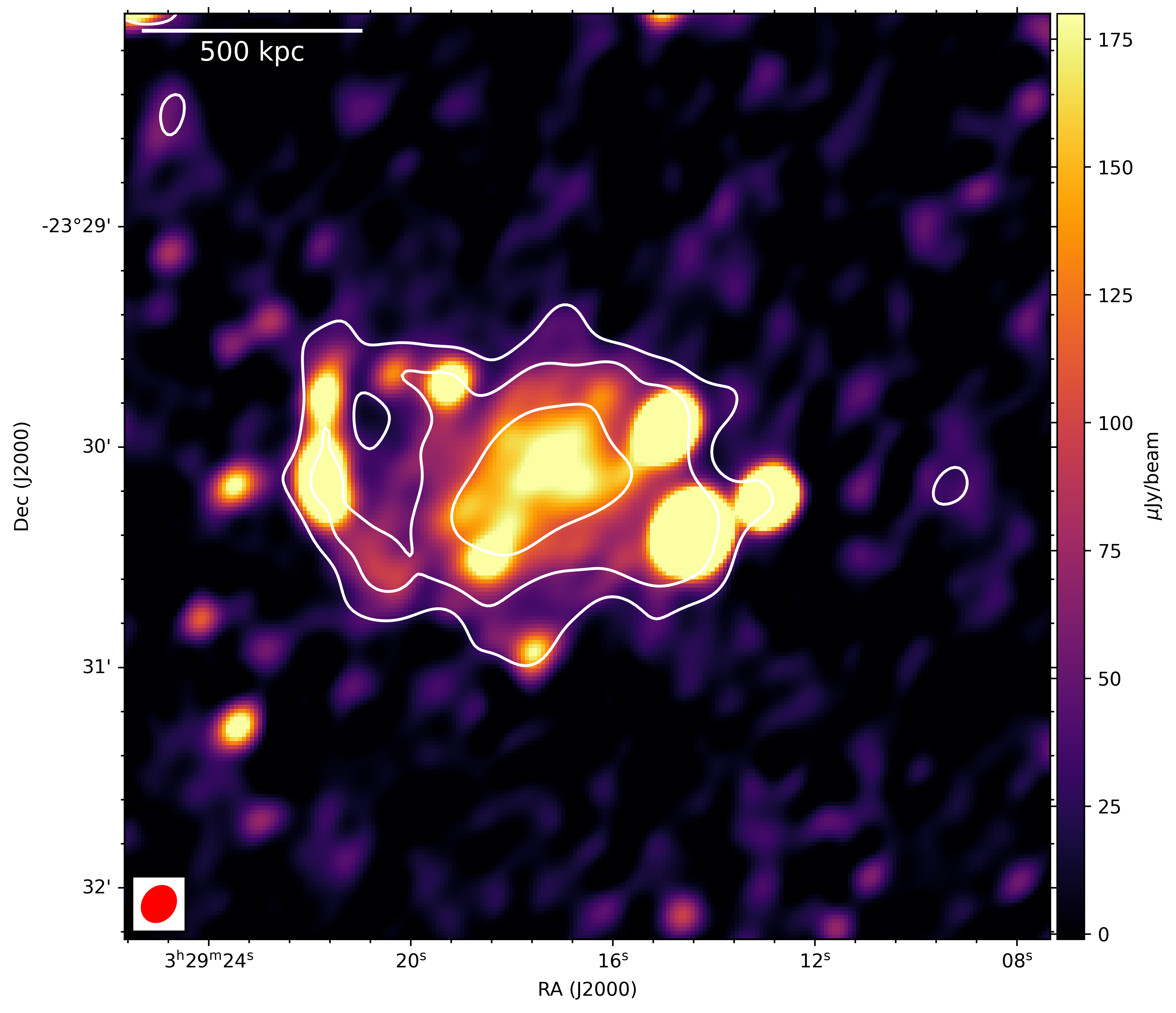}
    \caption{\textit{Left:}$L$-band full-resolution image of the radio halo. The local rms of the image is 1$\sigma$ = 6.6 $\upmu$Jy/beam; the beam size is indicated by the red ellipse. The overlaid contours are from the low-resolution image, which has a local rms of 1$\sigma$ = 7.1 $\upmu$Jy/beam. The contour levels are $\sigma \times$[3,6,10] \textit{Right:} $UHF$-band full-resolution image of the radio halo. The local rms of the image is 1$\sigma$ = 13.2 $\upmu$Jy/beam; the beam size is indicated by the red ellipse. The overlaid contours are from the low-resolution image, which has a local rms of 1$\sigma$ = 17.6 $\upmu$Jy/beam. The contour levels are $\sigma \times$[3,6,10]}
\label{fig:halo}
\end{figure*}

\par
\begin{figure}
	\includegraphics[width=\columnwidth]{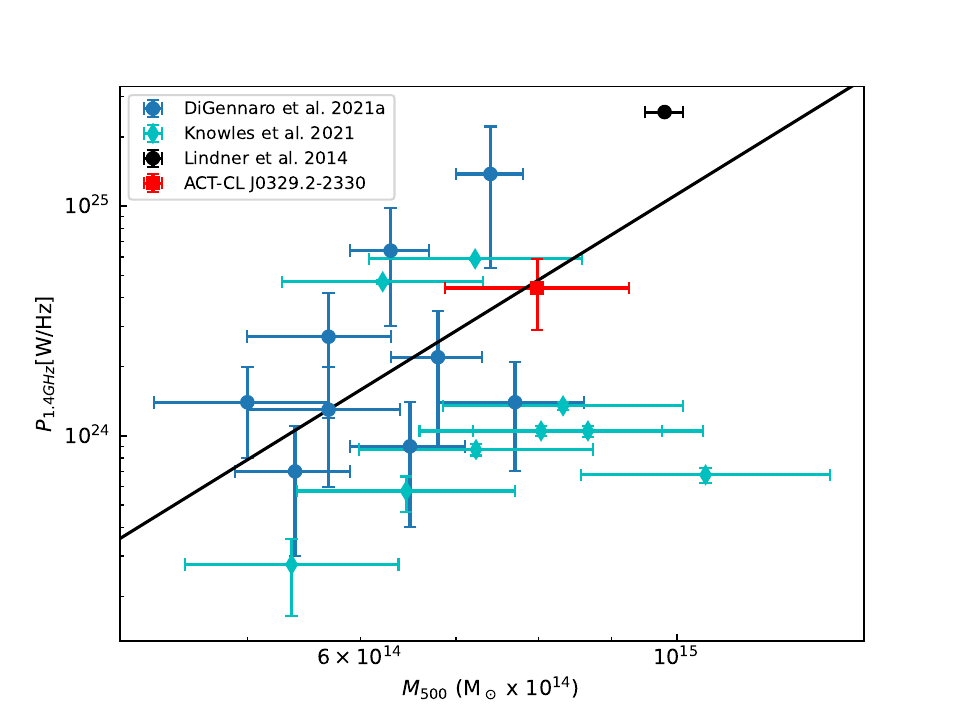}
    \caption{P$_\mathrm{1.4~GHz} - $M$_{500}$ correlation plot of radio halos. To account for the discrepancies between the ACT and \textit{Planck} mass derivations, we use the uncorrected M$_{500c}$ values for the ACT clusters \citep{2021ApJS..253....3H}. The solid line shows the best-fit relation, with parameters extracted from the low-redshift sample ($z\sim$0.2) presented in \citet{2013ApJ...777..141C}. The \citep{2021NatAs...5..268D} sample consists of radio halos at \textit{z} > 0.6, and the \citep{2021MNRAS.504.1749K} sample covers 0.22 < \textit{z} < 0.65. We also include El Gordo, which is at \textit{z} = 0.87 \citep{2014ApJ...786...49L}.}
\label{fig:halo-power}
\end{figure}

Using the 3$\sigma$ contours of the LR images (see Figure \ref{fig:halo}) as boundaries, we find that the radio halo has a linear scale of 1.1$\times$0.9\,Mpc at 1.28\,GHz and 900$\times$950\,kpc at 815.9\,MHz. The radio halo has a smooth, regular morphology that traces the thermal bremsstrahlung emission of the ICM, as seen in Figure \ref{fig:halo-xray}. We extract the flux densities in the same region; the radio halo has a flux density of 3.44 $\pm$ 0.2\,mJy and 6.11 $\pm$ 0.9\,mJy at $L$ and $UHF$-band, respectively. Using the multi-frequency flux density measurements, we derive the integrated radio halo spectral index to be $\alpha^{1.3GHz}_{0.8GHz}$ = 1.3 $\pm$ 0.4. The best-fit value of the integrated spectral index places the halo as a moderately steep spectrum radio halo. However, the large uncertainty associated with the spectral index limits our discussion. Bootstrapping the $L$-band flux density, we infer a halo radio power of P$_{\rm 1.4GHz} = (4.4 \pm 1.5)\times 10^{24}$\,W\,Hz$^{-1}$. The uncertainties associated with the integrated spectral index and the radio power are obtained from with 1000 Monte Carlo realisations. We compare this luminosity to the samples observed in \citet{2021NatAs...5..268D}, \citep{2021MNRAS.504.1749K}, and the radio halo in the El Gordo cluster \citep{2014ApJ...786...49L}. Plotting the P$_\mathrm{1.4~GHz} - $M$_{500}$ scaling relation, we find that the halo radio power falls within the scatter of both the high redshift and nearby radio halos. This is indicated in Figure \ref{fig:halo-power}, where the best-fit line is from scaling relation studies of nearby radio halos ($z \sim$0.2) carried out by \citet{2013ApJ...777..141C}. The radio luminosity of the halo falls within the correlation, indicating that the halo is as luminous as nearby ($z \sim$ 0.2) halos. This result further confirms that there is rapid magnetic field amplification in galaxy clusters at high redshifts, as discussed in \citet{2021NatAs...5..268D}.

\par
\begin{figure*}
    \centering
	\includegraphics[scale=0.55]{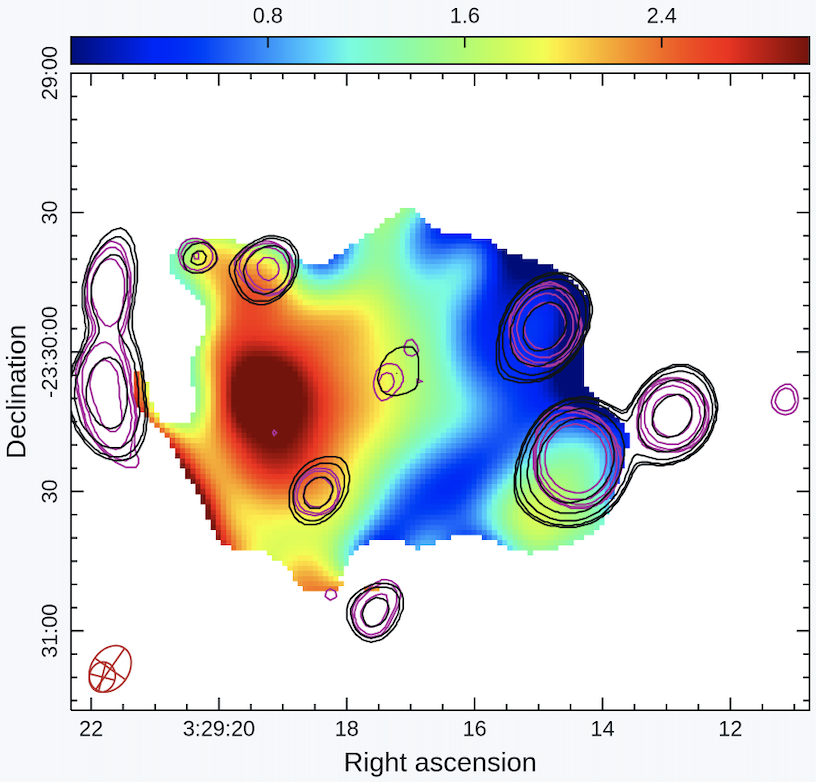}
    \includegraphics[scale=0.55]{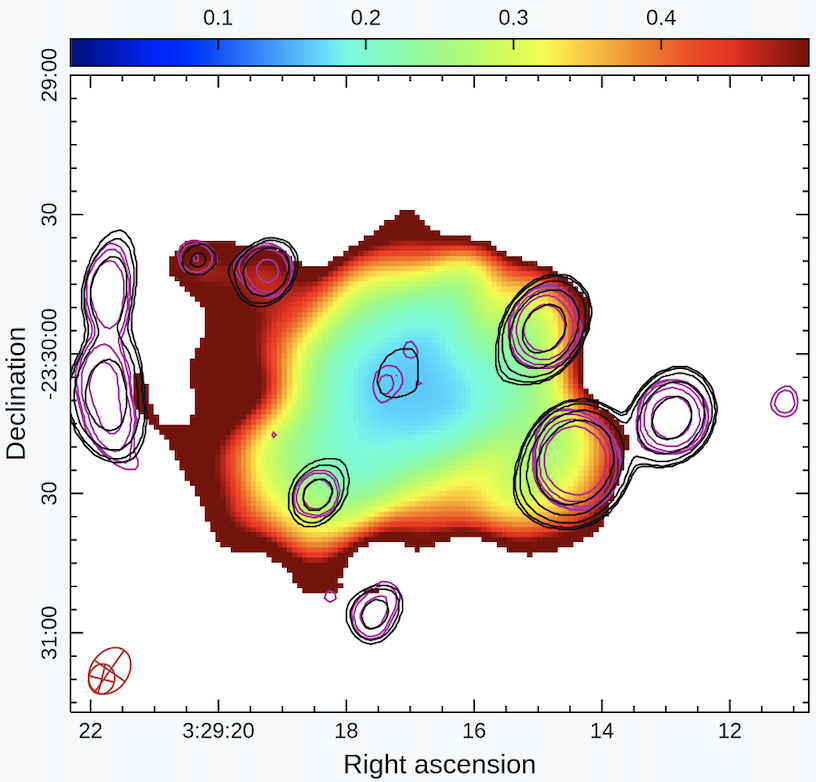}
    \caption{Spectral index map of the radio halo and the corresponding error map. The overlaid black and magenta contours are from the MeerKAT $UHF$-band and $L$-band's high-resolution images, respectively. For both images, the contour levels are $\sigma\times$[3,9,18,27]. 1$\sigma$ is 12.4\,$\upmu$Jy/beam for the $UHF$-band image and 4.9\,$\upmu$Jy/beam for the $L$-band image.}
\label{fig:halo-spec}
\end{figure*}
\begin{figure}
	\includegraphics[width=\columnwidth]{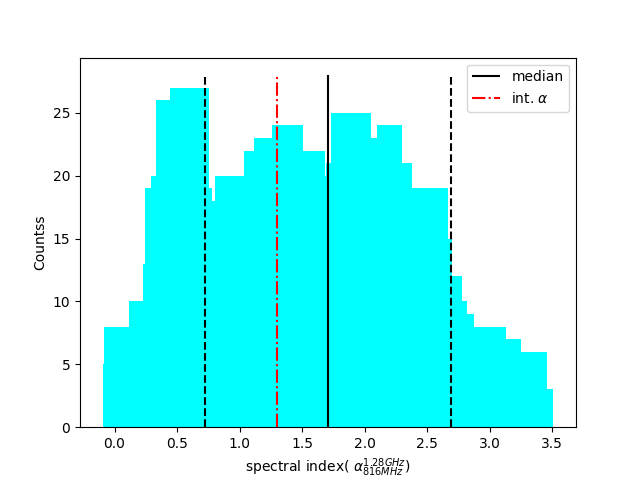}
    \caption{Histogram of the spectral index distribution for
the halo in ACT-CL\,J0329. The black solid line indicates the median spectral index value $\alpha$=1.7. The dashed black lines show the standard deviation around the
median, $\sigma$=0.98. The dot-dashed red line is the integrated spectral index.}
\label{fig:spix:hist}
\end{figure}

Spectral index studies of high redshift radio halos are vital for understanding the CRe re-acceleration mechanisms in the early universe. Figure \ref{fig:halo-spec} shows that the spectral index distribution varies from 0.5 to 3.3, with the eastern region of the halo having steeper spectra while the western region has flatter spectra. From the map, we extract a median spectral index of $\alpha$ = 1.7, which is steeper in comparison to the integrated spectral index value. The variations in spectral index maps can sometimes be due to measurement errors. According to \citet{2014A&A...561A..52V} and \citet{2017ApJ...845...81P}, if the variations are due to measurement errors, the median error value should be comparable to the standard deviation. The spectral index distributed for the maps is indicated in Figure \ref{fig:spix:hist}. The median error of the distribution is 0.05, and the standard deviation is 0.98, thus implying that the fluctuations in the spectral index map are intrinsic. A few studies have reported varying spectral index distribution in radio halos \citep{2020ApJ...897...93B,2021A&A...656A.154H,2023A&A...669A...1R}. According to \citet{2001MNRAS.320..365B} and \citet{2007MNRAS.378..245B}, fluctuations in the spectral index over the radio halo regions arise due to the non-uniformity
of the magnetic field distribution and different acceleration efficiencies. The fluctuations observed in the ACT-CL\,J0329 spectral index map are peculiar in that, the steeper spectral index values are concentrated in the eastern region, which does not overlap with a compact source. Thus, the physical interpretation could be that the turbulent energy is not homogeneously dissipated in the halo volume.

\section{Conclusion}
\label{sec:conclude}
In this letter, we have presented MeerKAT $L$ and $UHF$-band observations of ACT-CL\,J0329.2-2330, a galaxy cluster at $z$=1.23. The low-resolution images reveal a radio halo in the cluster; this is the highest redshift halo reported to date. The halo has an integrated spectral index of $\alpha$ = 1.3 $\pm$ 0.4 and a radio power of 4.4 $\pm$ 1.5$\times$10$^{24}$\,W\,Hz$^{-1}$. This indicates that the halo is as luminous as the halos found in nearby massive galaxy clusters. The spectral index map reveals varying spectral indices across the halo region.

\section*{Acknowledgements}
The MeerKAT telescope is operated by the South African Radio Astronomy Observatory, which is a facility of the National Research Foundation, an agency of the Department of Science and Innovation. We acknowledge the assistance of the South African Radio Astronomy Observatory (SARAO) science commissioning and operations team (led by Sharmila Goedhart). We acknowledge the financial assistance of the SARAO towards this research. KM and MH acknowledge support from the National Research Foundation of South Africa. CS acknowledges support from the Agencia Nacional de Investigaci\'on y Desarrollo (ANID) through Basal project FB210003.



\section*{Data Availability}
The data derived in this study are available upon reasonable request from the corresponding author.


\bibliographystyle{aa}
\bibliography{mkat_highz} 



\appendix

\section{Ancillary images of ACT-CL\,J0329.2-2330}

\begin{figure}
   \centering
	\includegraphics[width=0.5\textwidth]{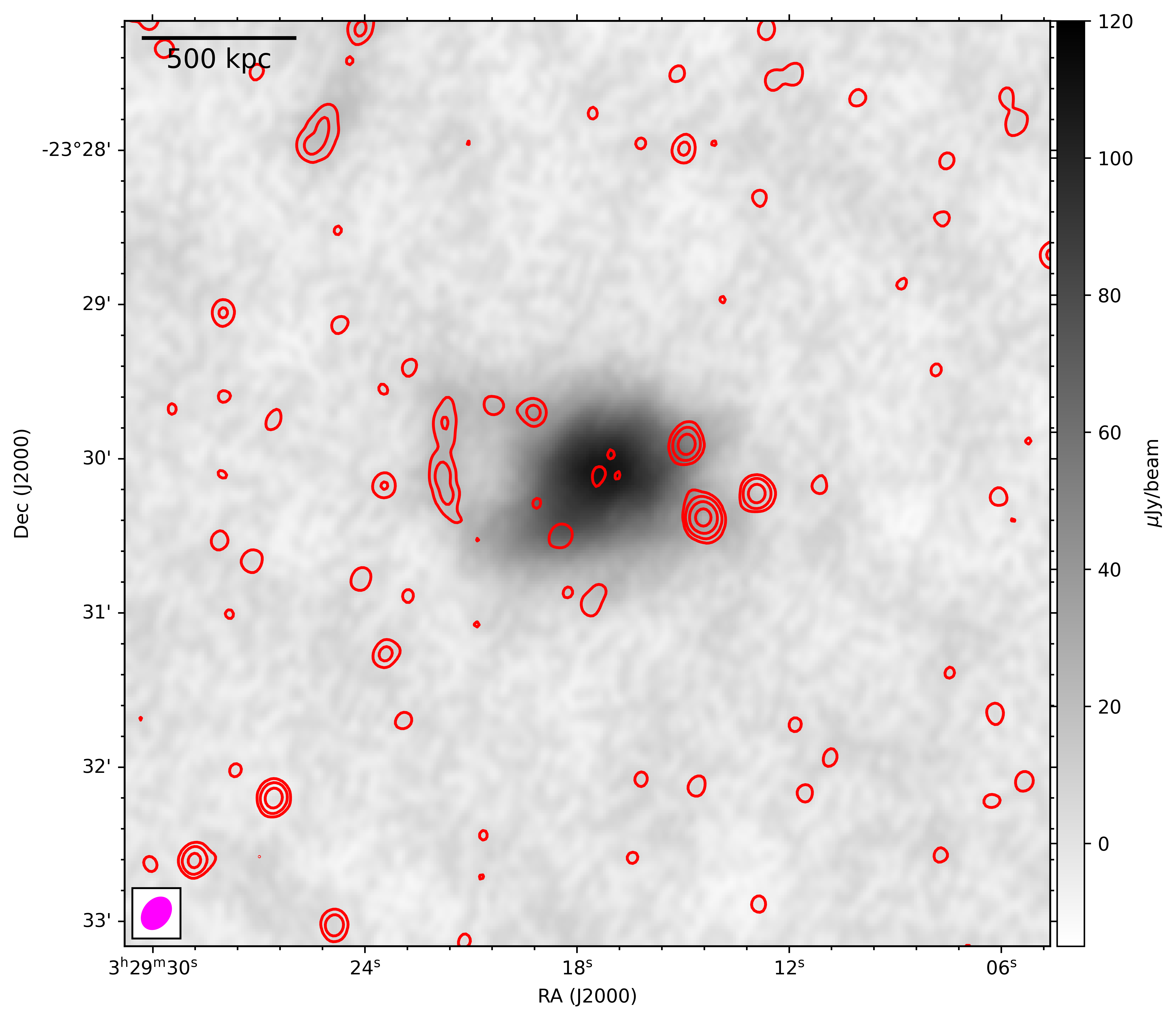}
	\caption{The L-band full-resolution compact-source-subtracted image with contour overlays of the high-resolution image. The contour levels are $\sigma\times$[3,12,24,30], where 1$\sigma$ is 4.9 $\upmu$Jy/beam. The beam of the high-resolution image is 6.5$\arcsec\times$5.5$\arcsec$, 166$\degree$ pa.}
    \label{fig:l-subtracted}
\end{figure}

\begin{figure}
   \centering
	\includegraphics[width=0.5\textwidth]{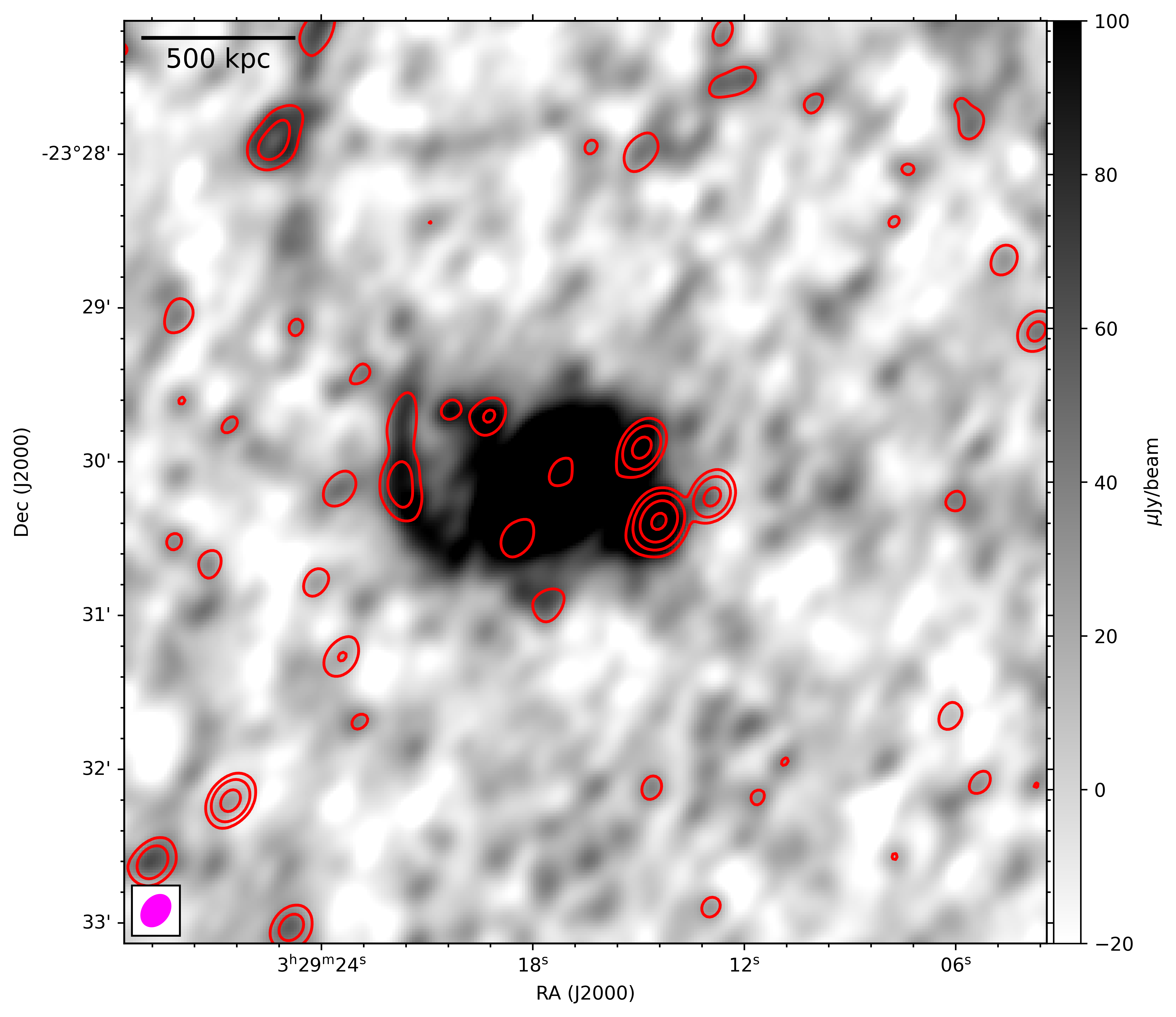}
	\caption{The UHF-band full-resolution compact-source-subtracted image with contour overlays of the high-resolution image. The contour levels are $\sigma\times$[3,12,24,30], where 1$\sigma$ is 12.4 $\upmu$Jy/beam. The beam of the high-resolution image is 10.7$\arcsec\times$8.1$\arcsec$, 146 $\degree$ pa.}
    \label{fig:uhf-subtracted}
\end{figure}

\begin{figure}
   \centering
	\includegraphics[width=0.5\textwidth]{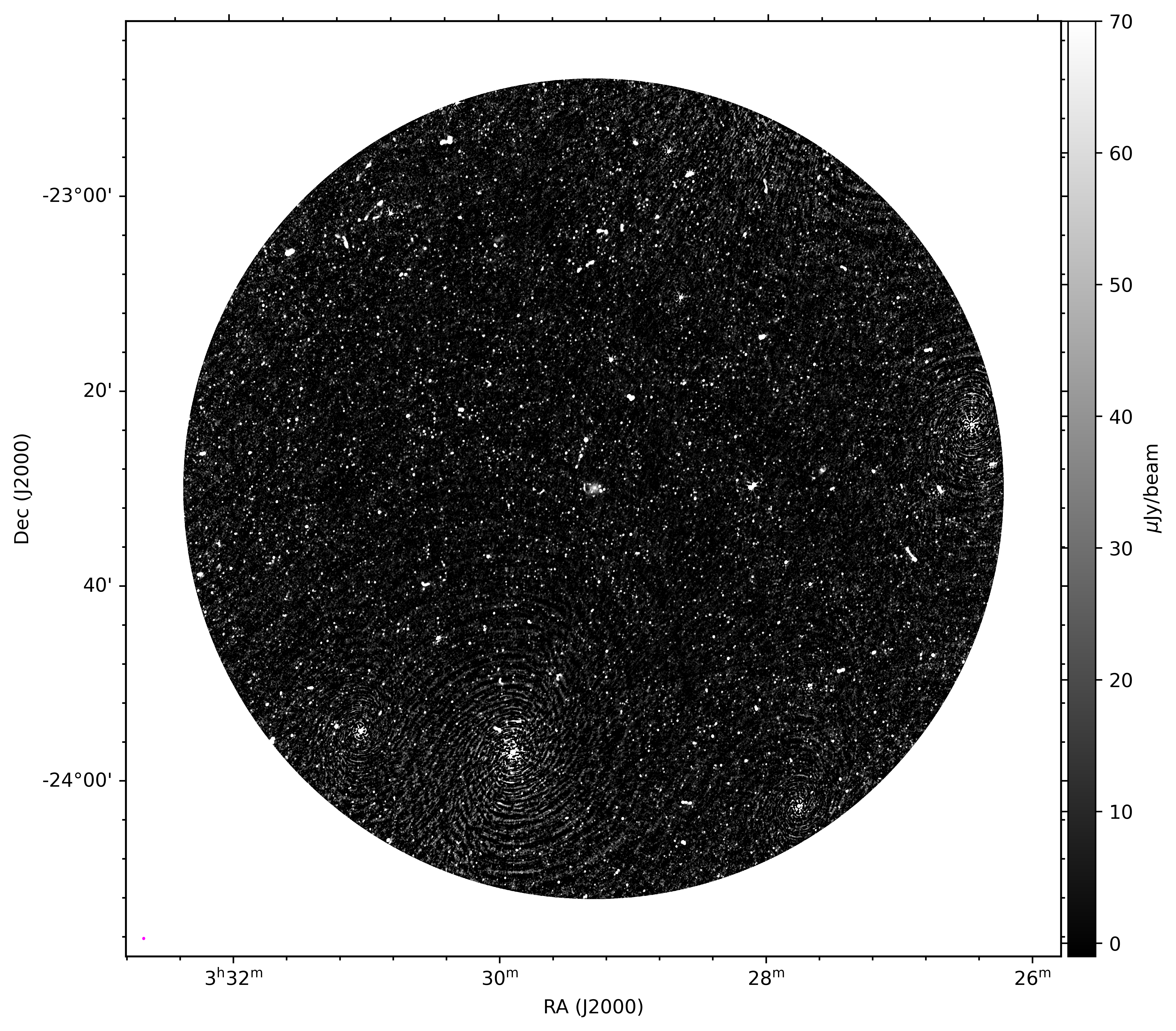}
	\caption{ The MeerKAT L-band full field of view image of ACT-CL\,J0329. The rms noise of the image is 6.6 $\upmu$Jy/beam.}
    \label{fig:l-fullfield}
\end{figure}

\begin{figure}
   \centering
	\includegraphics[width=0.5\textwidth]{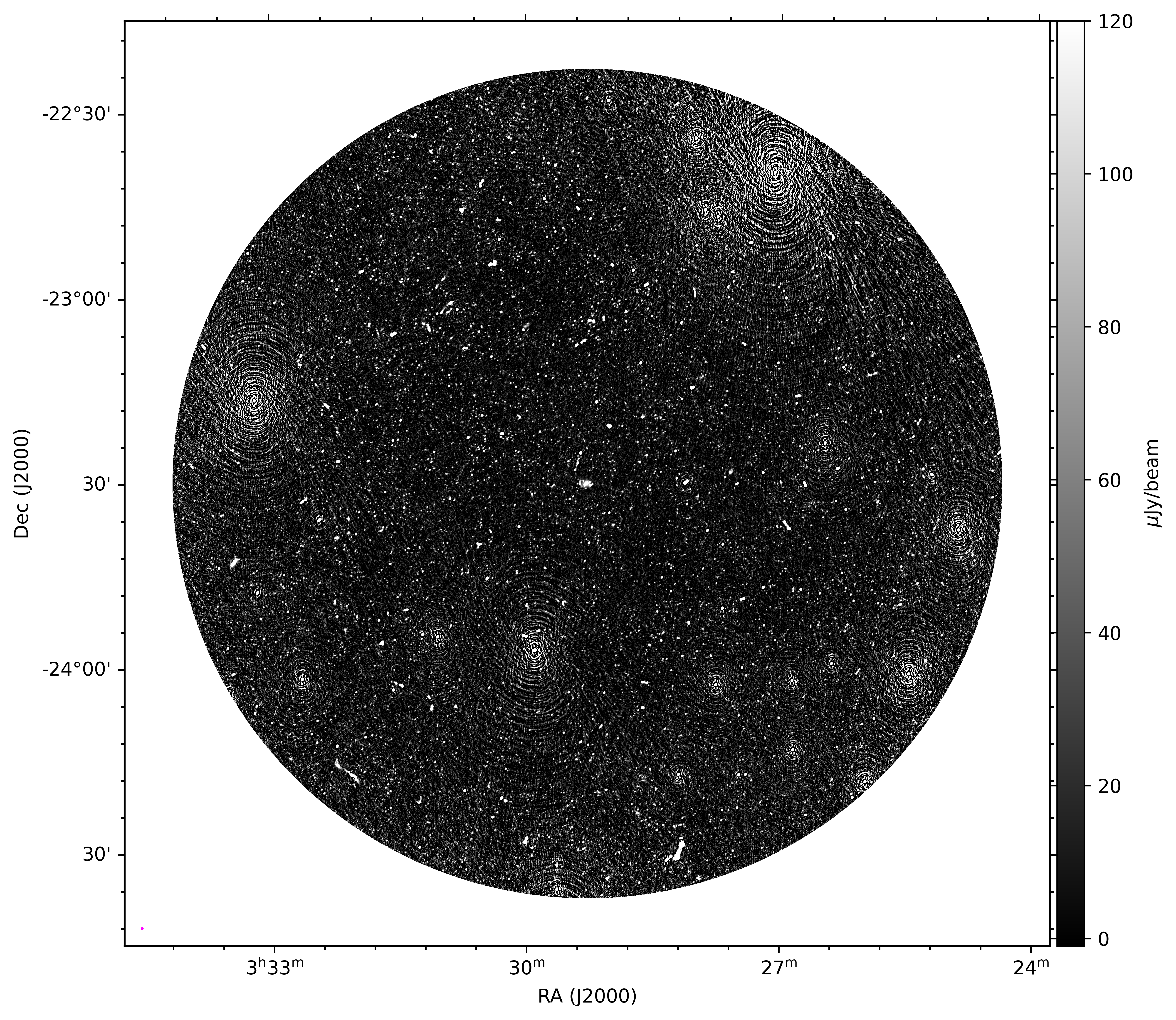}
	\caption{ The MeerKAT UHF-band full field of view image of ACT-CL\,J0329. The rms noise of the image is 13.2 $\upmu$Jy/beam.}
    \label{fig:uhf-fullfield}
\end{figure}

\begin{figure}
   \centering
	\includegraphics[width=0.5\textwidth]{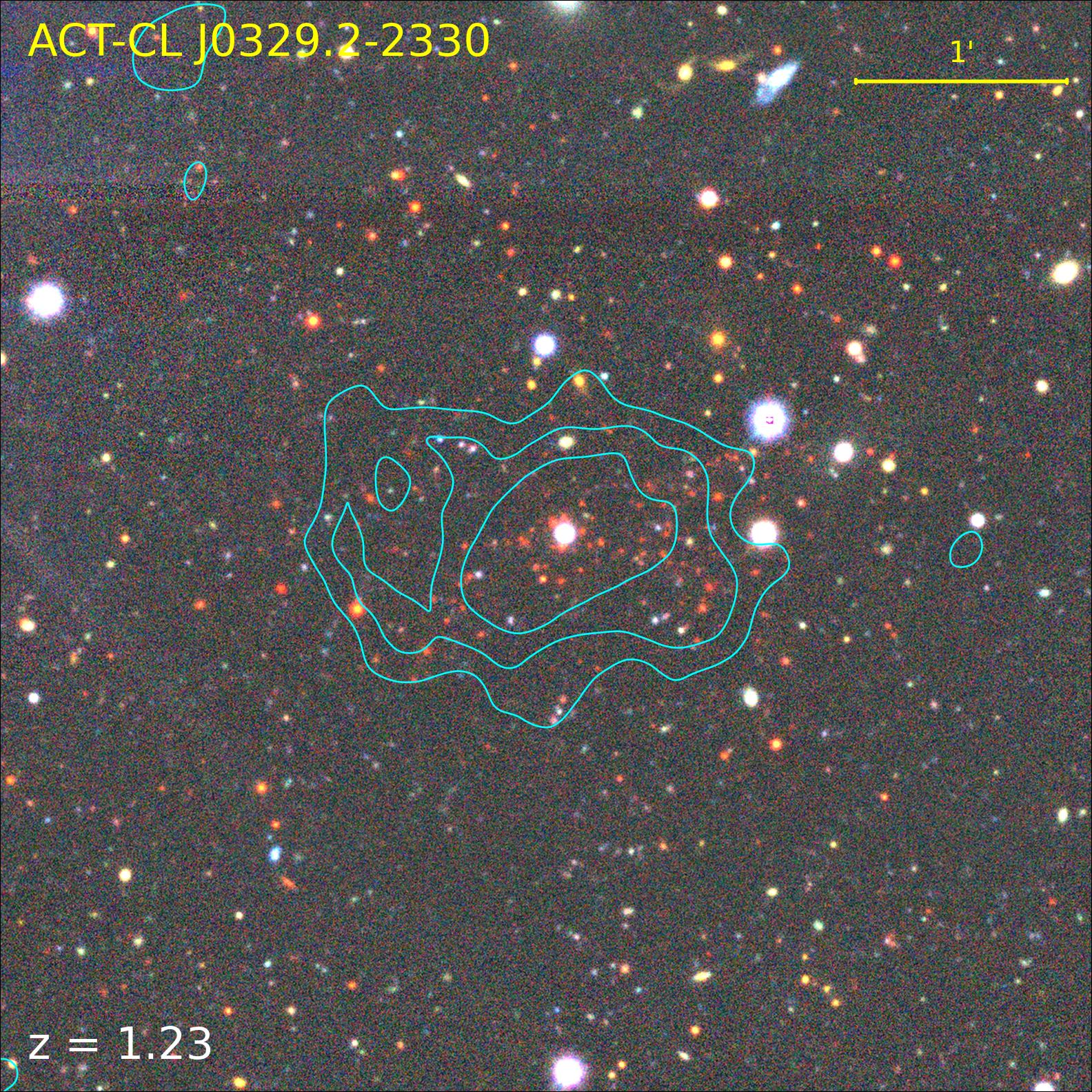}
	\caption{Optical \textit{grz} image of ACT-CL\,J0329 from DECaLS \citep{Dey_2019} The overlaid contours are from the MeerKAT $UHF$-band LR images, the contour levels are $\sigma \times$[3,6,10]; where 1$\sigma$ = 17.6 $\upmu$Jy/beam. The white source at the centre is a foreground star located at a distance of $\approx 1500$\,pc as recorded in Gaia DR3 \citep{Gaia_2023}.}
    \label{fig:optical}
\end{figure}


\label{lastpage}
\end{document}